\documentclass[onecolumn,usenatbib,useAMS]{mnras}

\usepackage{amsbsy,amssymb,amsmath,bm,eqnarray}
\usepackage{graphicx,subfigure}
\usepackage{lineno,color}

\title[Why hot Jupiters can be large but not too large]{Why hot Jupiters can be large but not too large}
\author[Qiang Hou and Xing Wei]{Qiang Hou and Xing Wei$^*$ \\ Department of Astronomy, Beijing Normal University, China \\ $*$ corresponding author: xingwei@bnu.edu.cn}

\begin{document}
\label{firstpage}
\maketitle

\begin{abstract}
Tidal heating is often used to interpret ``radius anomaly'' of hot Jupiters (i.e. radii of a large fraction of hot Jupiters are in excess of 1.2 Jupiter radius which cannot be interpreted by the standard theory of planetary evolution). In this paper we find that tidal heating induces another phenomenon ``runaway inflation'' (i.e. planet inflation becomes unstable and out of control when tidal heating rate is above its critical value). With sufficiently strong tidal heating, luminosity initially increases with inflation, but across its peak it decreases with inflation such that heating is stronger than cooling and runaway inflation occurs. In this mechanism, the opacity near radiative-convective boundary (RCB) scales approximately as temperature to the fourth power and heat cannot efficiently radiate away from planet interior, which induces runaway inflation (similar to a tight lid on a boiling pot). Based on this mechanism, we find that radii of hot Jupiters cannot exceed $2.2R_J$, which is in good agreement with the observations. We also give an upper limit for orbital eccentricity of hot Jupiters. Moreover, by comparison to the observations we infer that tidal heating locates near RCB.
\end{abstract}

\section{Introduction}\label{sec:introduction}

\begin{figure}
\centering
\subfigure[]{\includegraphics[width=0.6\textwidth]{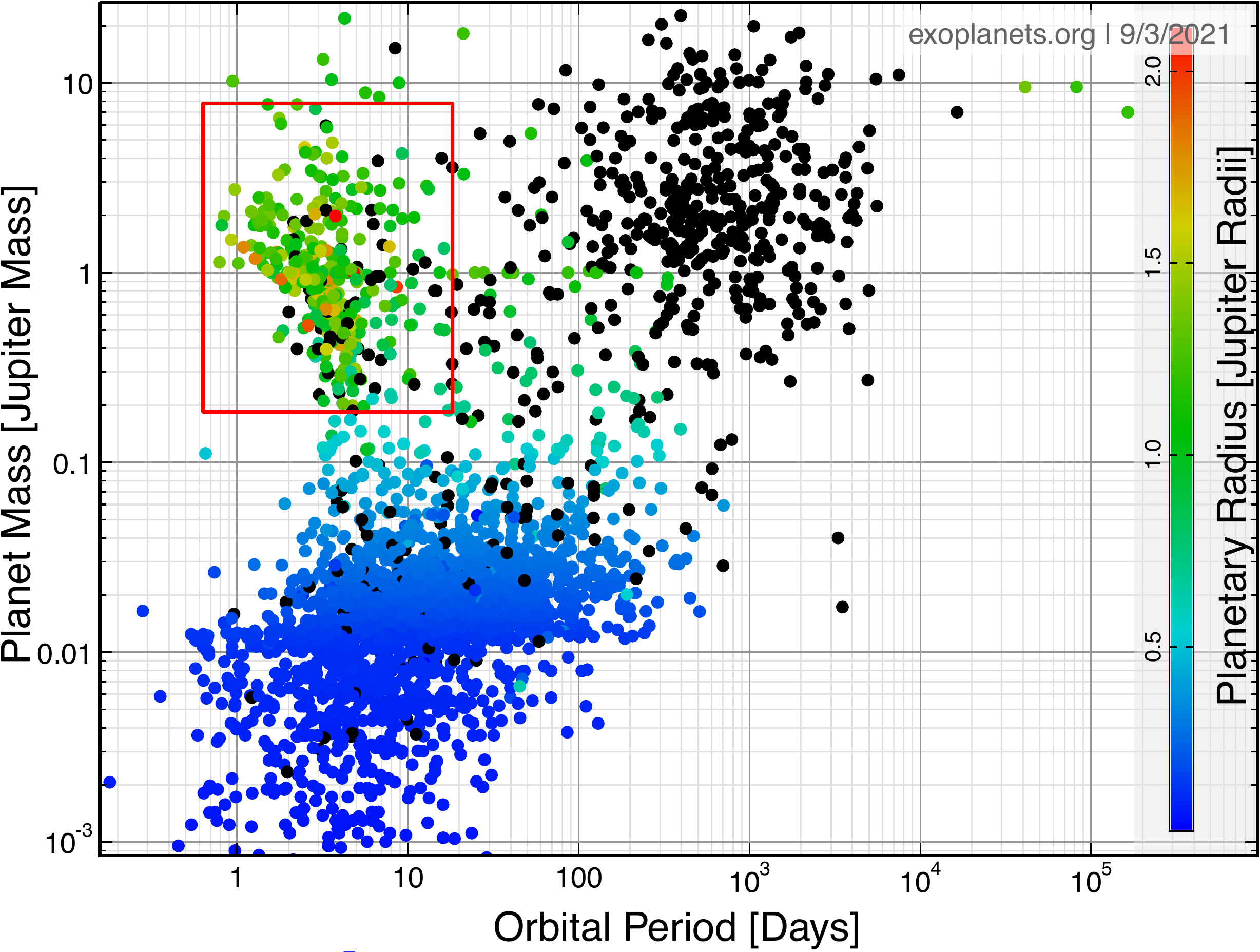}\label{fig1a}}
\subfigure[]{\includegraphics[width=0.35\textwidth]{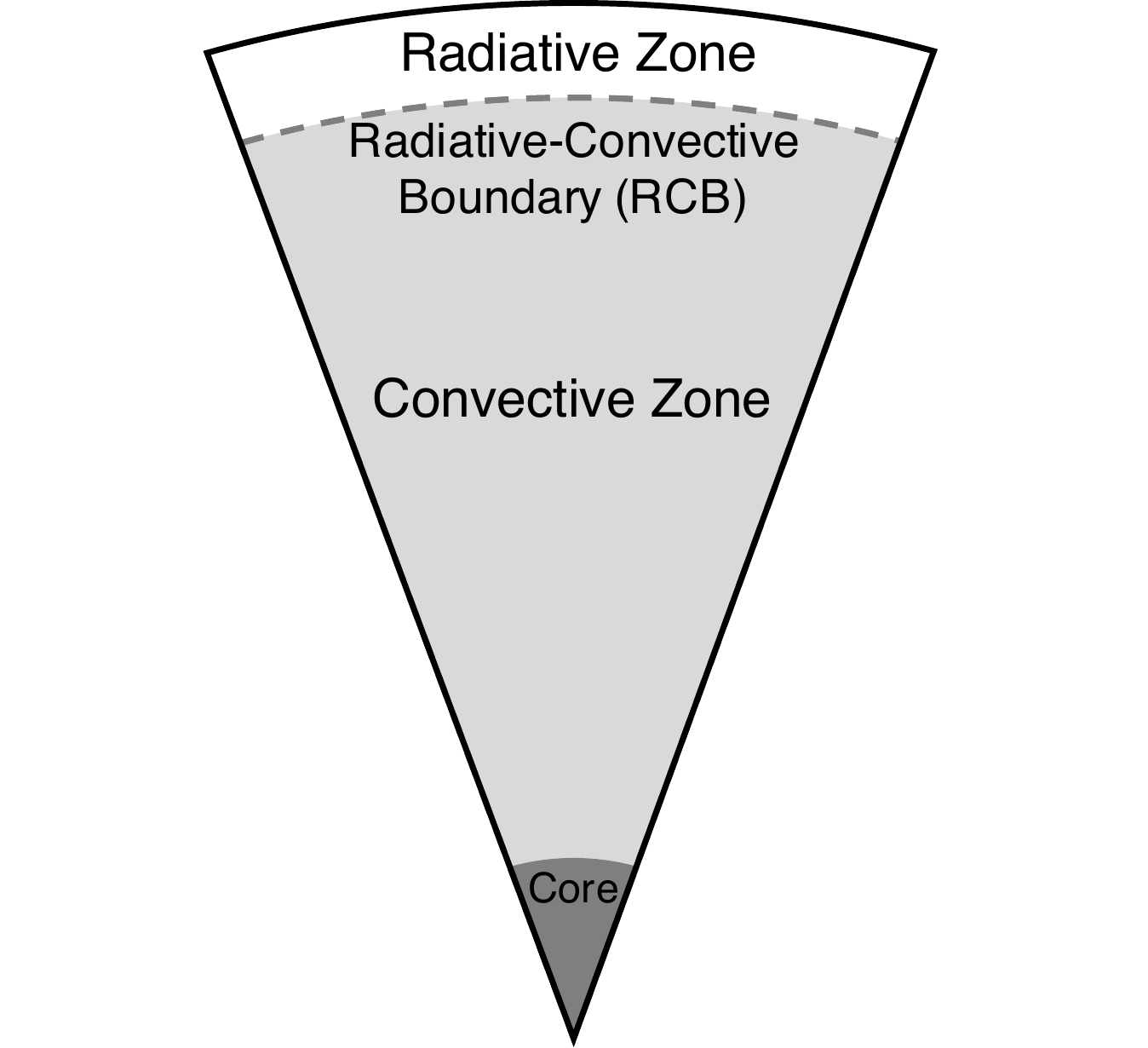}\label{fig1b}}
\caption{(a) Confirmed exoplanets (http://exoplanets.org/), in the red rectangle are ``hot Jupiters'', and black points denote planets without known radii. (b) The sketch of planet interior.}
\end{figure}

More than 4000 extrasolar planets (i.e. exoplanets) have been detected as shown in Figure \ref{fig1a}. The red rectangle shows ``hot Jupiters'' with their mass and radius comparable to those of Jupiter but on a close orbit with period $\lesssim 10$ days. A large fraction of hot Jupiters possess radii in excess of $1.2R_J$ ($R_J$ being Jupiter radius), which cannot be interpreted by the standard theory of planetary evolution \citep{stevenson1977, fortney2007}, no matter whether the initial entropy is high or low \citep{marley2007}. This is a long-standing problem called ``radius anomaly'', and some interpretations have been proposed, e.g. \citet{Bodenheimer2001, Gu2003, Gu2004, lin2008, Batygin2010, youdin2010, Guillochon2011, Burrows2013, Komacek2017}. In these interpretations the physical mechanisms involve tidal heating, Ohmic heating, inward transfer of irradiation, and high opacity. The former three mechanisms attribute to the energy sources and the latter one to the suppression of radiative cooling.

However, up to now, exoplanets with radius $>2.2R_J$ have not been detected except perhaps PDS 70 b \citep{wang2020}, a newly born planet in a protoplanetary disk (the error bars in \citep{wang2020} do not rule out $<2.2R_J$). In this paper we will study why hot Jupiters can be large but not too large. We find that the reason is the runaway inflation due to tidal heating combined with high opacity near radiative-convective boundary (RCB). That is, when tidal heating rate is above its critical value, luminosity will decrease with inflation such that inflation becomes unstable and out of control. This runaway inflation mechanism has never been reported in the previous studies of planet inflation. In \S\ref{sec:methods} we show the mathematical equations for planetary evolution and the numerical method to solve them. In \S\ref{sec:inflation} we choose a set of typical parameters to show runaway inflation and give the physical interpretation. In \S\ref{sec:parameters} we investigate the influence of parameters on runaway inflation. In \S\ref{sec:observation} we compare our theoretical predictions to the observations. In \S\ref{sec:summary} we give a brief summary.

\section{Methods} \label{sec:methods}

The sketch of planet interior is shown in Figure \ref{fig1b}. The mass of thin radiative zone is less than $1\%$ of the total mass. The convective zone is thick and considered to be adiabatic because heat transfer is very fast and efficient via convection. The interface between radiative and convective zones is called radiative-convective boundary (i.e. RCB). A rocky core lies near the center as core accretion model describes \citep{Safronov1972, Mizuno1978, Mizuno1980, Pollack1996}.

Equations for planetary evolution are similar to those of stellar evolution in the absence of nuclear energy \citep{kippenhahn1990},
\begin{equation}\label{evolution}
\begin{aligned}
&\frac{\partial r}{\partial m}=\frac{1}{4 \pi r^{2} \rho}, \hspace{3mm} 
\frac{\partial P}{\partial m}=-\frac{G m}{4 \pi r^{4}}, \hspace{3mm} 
\frac{\partial T}{\partial m}=-\frac{G m T}{4 \pi r^{4} P} \nabla, \\
& \nabla=\left\{
\begin{aligned}
& \nabla_{\rm ad}=\left(\frac{\partial\ln T}{\partial\ln P}\right)_s \hspace{5mm}\mbox{in convective zone}, \\
& \nabla_{\rm rad}=\frac{3}{16 \pi a c G} \frac{\kappa \mathcal{L} P}{m T^{4}} \hspace{5mm}\mbox{in radiative zone},
\end{aligned} 
\right. \\
&\frac{\partial \mathcal{L}}{\partial m}=-c_p\frac{\partial T}{\partial t}+\frac{\delta}{\rho}\frac{\partial P}{\partial t}+\varepsilon(m), \hspace{3mm}
\delta=-\left(\frac{\partial\ln\rho}{\partial\ln T}\right)_P.
\end{aligned}
\end{equation}
In \eqref{evolution} we use the conventional notation for all the variables as in \citep{kippenhahn1990}. Luminosity at surface
\begin{equation}
\mathcal L_{intr}=4 \pi \sigma R_{p}^{2}\left(T_{surf}^{4}-T_{e q}^{4}\right)
\end{equation}
is the intrinsic luminosity arising from planet interior. The equilibrium temperature $T_{e q}$ measures irradiation from host star. $\varepsilon(m)$ is the rate of energy per unit mass injected into planet, e.g. tidal heating. $\varepsilon(m)$ is a function of mass coordinate because tidal heating is presumably not uniform. Turbulent dissipation of equilibrium tides is strong near top of convective zone because of low density $\rho$ (note that turbulent viscosity is proportional to convective velocity and the latter is proportional $\rho^{-1/3}$), and waves of dynamical tide tend to nonlinearly break and deposit energy near top of convective zone \citep{goodman1996,wu2018}. Thus, $\varepsilon(m)$ due to tidal heating is assumed to obey Gaussian distribution
\begin{equation}
\varepsilon(m)=\frac{\varepsilon_{inj}}{\sigma\sqrt{2 \pi} } \exp \left[-\frac{(m-m_{0})^{2}}{2 \sigma^{2}}\right].
\end{equation}
$m_0=0.99M_p$ denotes the location of RCB. We keep the rate of total energy injected into planet $\varepsilon_{inj}=\int_{M_{core}}^{M_{p}}\varepsilon(m)dm$ constant. Different values of mass coordinate $m_0$ will be tested with the fixed $\sigma=0.05M_p$. Uniform distributions will be also tested for comparison. 

We use \texttt{MESA} code \citep{Paxton2011,Paxton2013,Paxton2015,Paxton2018,Paxton2019} to solve Equations \eqref{evolution} along with the equation of state from \citet{Saumon1995} and the tabulated opacity from \citet{Freedman2008}. The element abundances are hydrogen $X=0.74$, helium $Y=0.24$ and metal $Z=0.02$. The initial age 3 Myr is set, and the initial radius $\approx 1.9R_{J}$ is used \citep{Pollack1996}. For the inner boundary, a coreless planet and a planet with a rocky core of density $\rho_{\rm core}=5 \;{\rm g/cm^3}$ will be both considered. For the outer boundary, we impose equilibrium temperature $T_{eq}\approx 1500$ K, corresponding to a planet with period $\approx 3$ days around a solar-type star.

\section{Runaway inflation}\label{sec:inflation}

\begin{figure}
\centering
\subfigure[Radius evolution]{\includegraphics[width=0.32\textwidth]{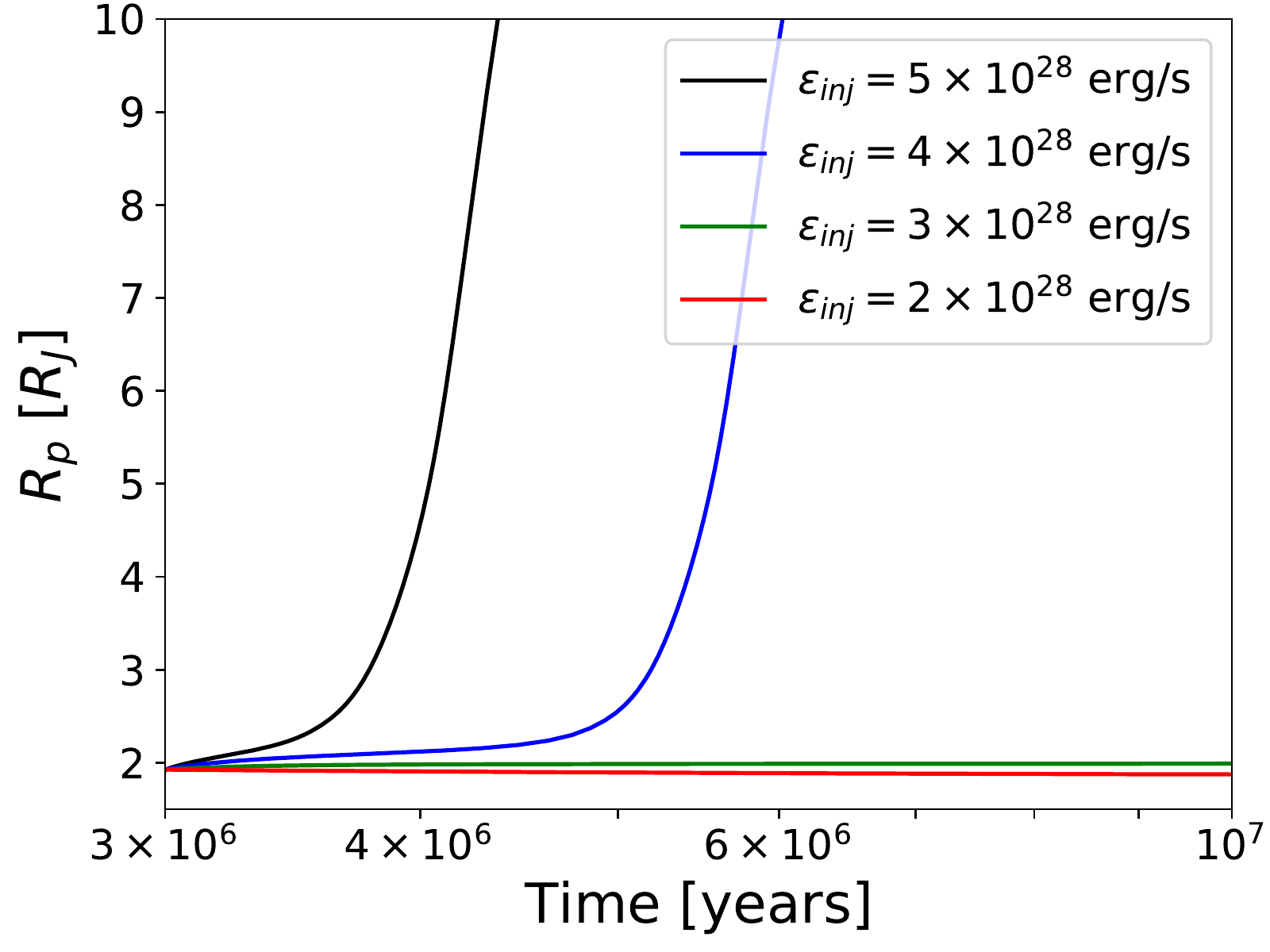}\label{fig2a}}
\subfigure[Intrinsic luminosity versus radius]{\includegraphics[width=0.32\textwidth]{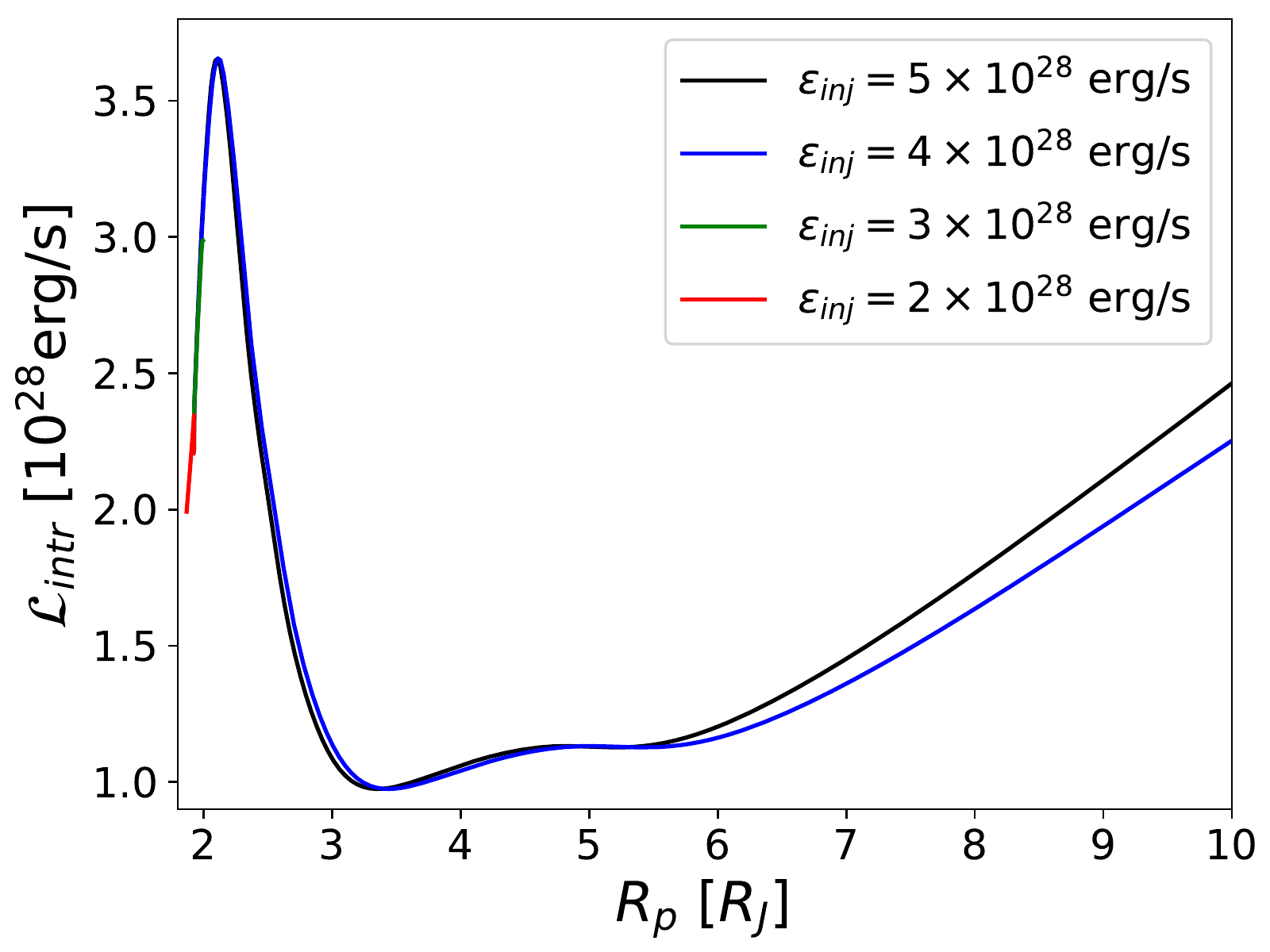}\label{fig2b}} \\
\subfigure[Luminosity at RCB versus temperature (dot: runaway point, dashed line: fitting for post-runaway inflation)]{\includegraphics[width=0.32\textwidth]{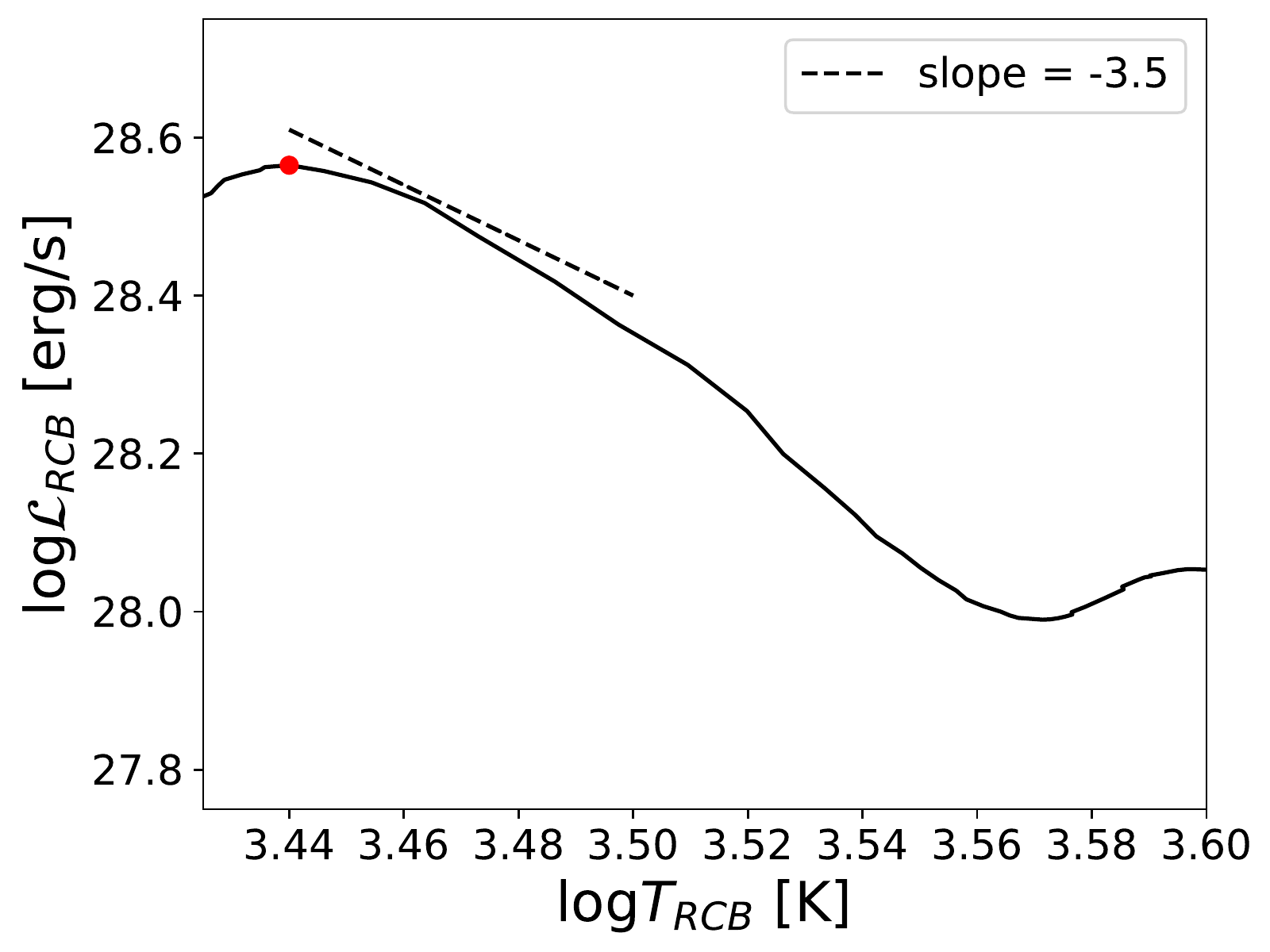}\label{fig2c}}
\subfigure[Test of opacity to illustrate the physical mechanism of runaway inflation]{\includegraphics[width=0.32\textwidth]{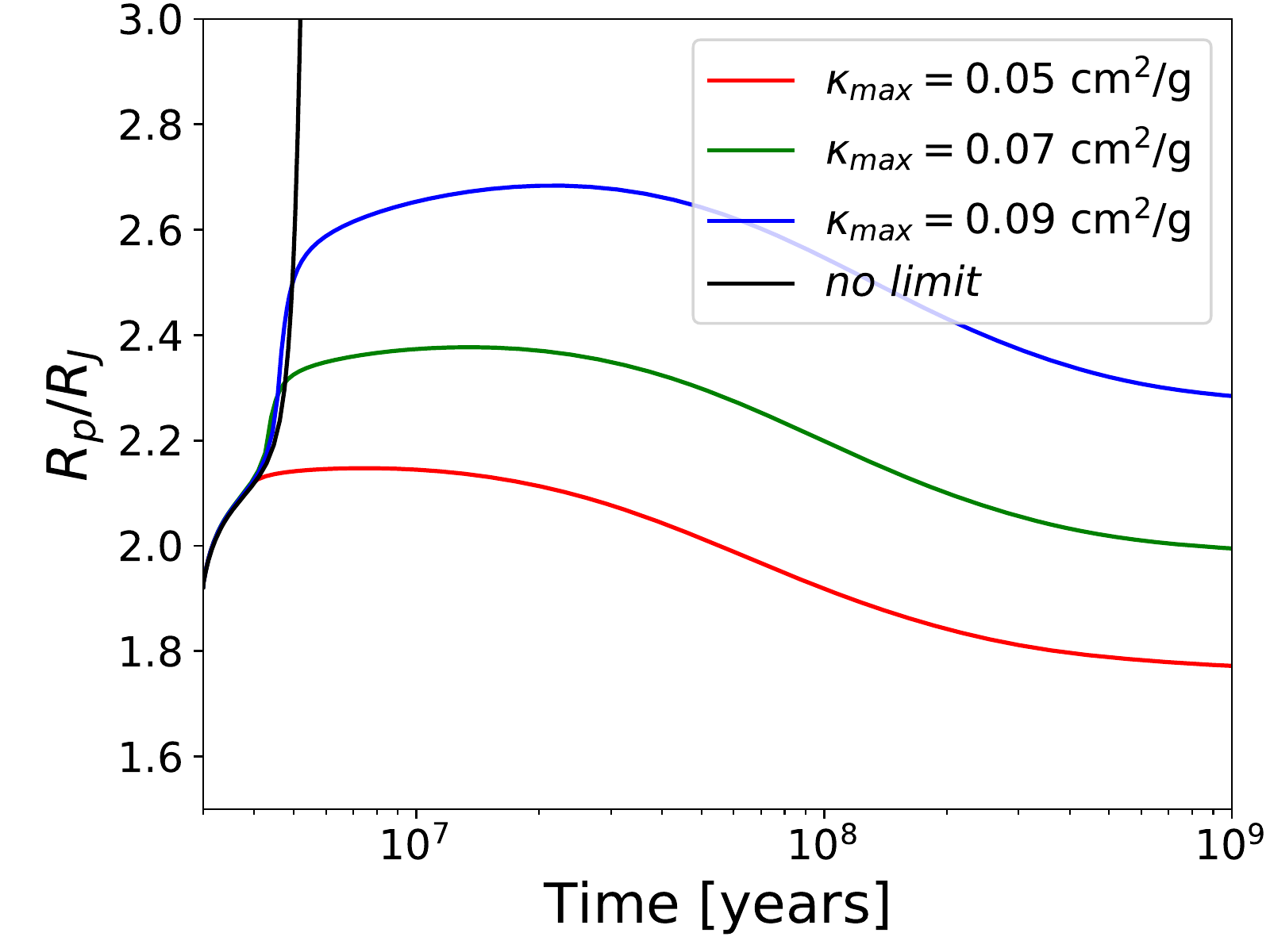}\label{fig2d}}
\subfigure[Pressure-density relation to fit with polytropic models (dots: RCB)]{\includegraphics[width=0.32\textwidth]{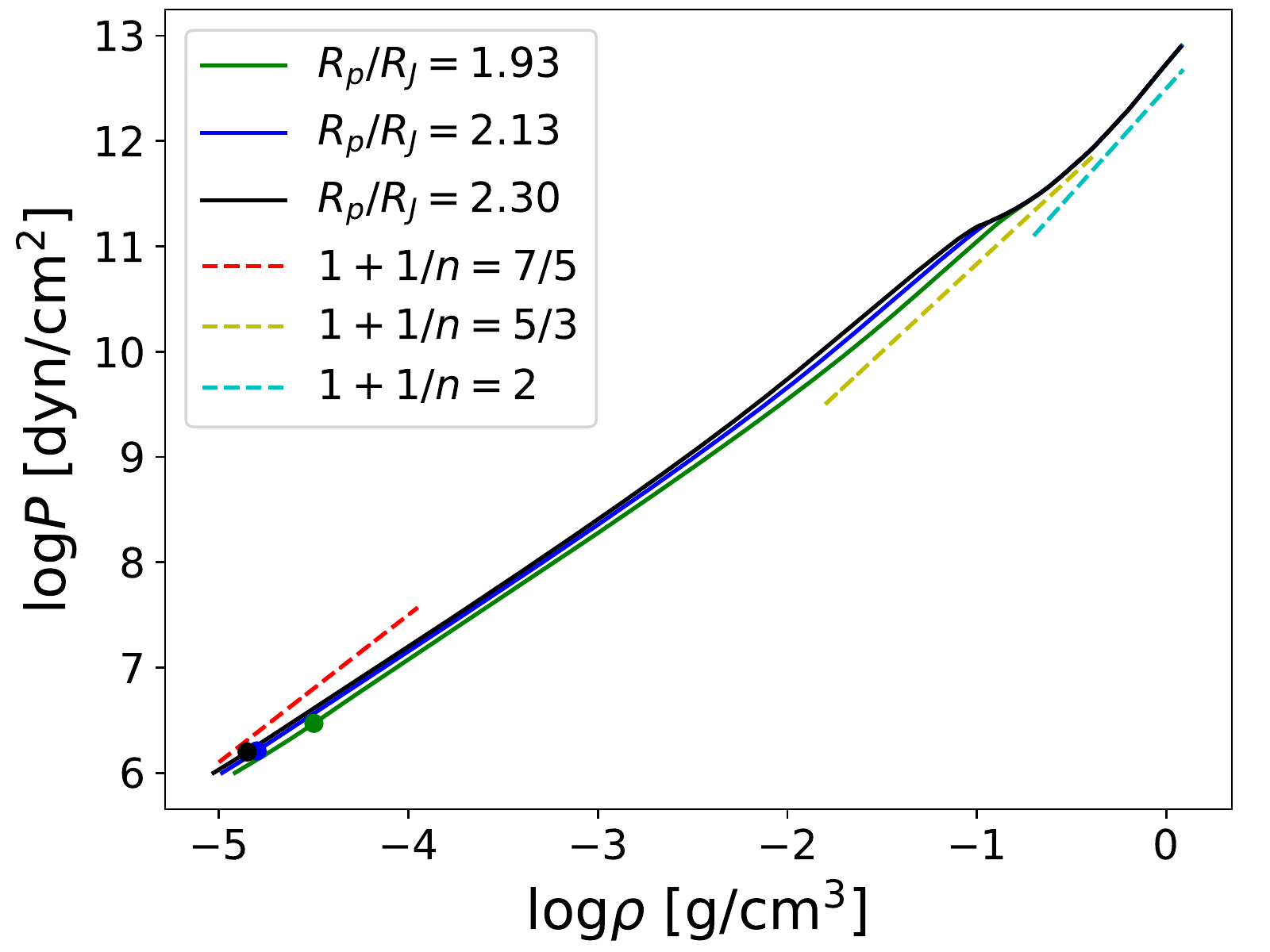}\label{fig2e}}
\caption{Runaway inflation. Subfigures (a) and (b) show the phenomenon of runaway inflation, and subfigures (c), (d) and (e) illustrate the physical mechanism of runaway inflation.}
\end{figure}

We choose typical parameters, namely $M_p=M_J$, $M_{core}=10M_\oplus$, $T_{eq}=1500~K$ and $m_0=0.99M_p$ (i.e. tidal heating near RCB), to test the total energy rate $\varepsilon_{inj}$. Figure \ref{fig2a} shows the radius evolution. With low $\varepsilon_{inj}$ (red and green curves) planet does not inflate too much because injected energy is balanced by luminosity radiated away. However, with high $\varepsilon_{inj}$ (blue and black curves) planet smoothly inflates to $2.13R_J$ and then rapidly inflates beyond Hill radius $(M_\odot/M_J)^{1/3}R_J\approx 10 R_J$ to be destroyed by its host star. This unstable inflation out of control is called {\bf runaway inflation} due to energy imbalance (i.e. tidal heating is stronger than radiative cooling). Higher $\varepsilon_{inj}$ corresponds to earlier runaway inflation (comparison between blue and black curves). Figure \ref{fig2b} shows luminosity-radius relation. With low $\varepsilon_{inj}$ (red and green curves) luminosity initially increases with inflation until it balances with injected energy, such that inflation stops at $\approx 2R_J$. However, with high $\varepsilon_{inj}$ (blue and black curves) luminosity continues to increase to a peak called runaway point. At runaway point, the corresponding luminosity is called {\bf critical rate of injected energy} $\varepsilon_{cri}=3.67\times10^{28}\mathrm{~erg/s}$ and the corresponding radius is called {\bf runaway radius} $R_{run}=2.13R_J$. Across runaway point, luminosity decreases with inflation and energy imbalance takes place, i.e. energy injected into planet cannot efficiently radiate away from planet interior. Consequently, inflation becomes unstable and runaway inflation occurs.

To understand the physics of runaway inflation, we need to make clear how luminosity changes at runaway point. Before runaway inflation occurs, luminosity increases smoothly with inflation and temperature also increases. When temperature near RCB reaches $\approx 3000$ K, luminosity near RCB turns to be heavily suppressed. We now simply analyze luminosity-temperature relation near RCB. In Equations \eqref{evolution} $\nabla_{\rm rad}$ is continuous across RCB and equal to $\nabla_{\rm ad}$ which is independent of temperature, such that we readily find $\mathcal L_{RCB}\propto T^4/(P\kappa)$. With the help of polytrope model $P\propto \rho^{1+1/n}$ where $n$ is polytropic index, we obtain $P\propto T^{n+1}$ for ideal gas. We assume that opacity obeys the power law $\kappa\propto T^{\alpha}$. Thus, we are led to $\mathcal L_{RCB}\propto T^{3-n-\alpha}$. Inserting $n=2.5$ of diatomic molecular hydrogen and $\alpha\approx 4$ due to the contribution from some molecules at temperature $\approx 3000$ K near RCB \citep{Freedman2008}, we find $\mathcal L_{RCB}\propto T^{-3.5}$. Figure \ref{fig2c} shows luminosity-temperature relation at RCB (red dot denotes runaway point) and it verifies our power law $\mathcal L_{RCB}\propto T^{-3.5}$ when runaway inflation occurs. That is, a small increase in temperature near RCB will induce a strong suppression on luminosity, such that heat cannot efficiently radiate away and planet will inflate out of control, namely runaway inflation. This process is similar to a tight lid on a boiling pot. Runaway inflation is essentially caused by the suppression of radiation near RCB (i.e. $\kappa\propto T^4$).

To validate our interpretation, we carry out numerical tests by setting the upper limit for opacity. Figure \ref{fig2d} shows radius evolution with $\varepsilon_{inj}=4 \times 10^{28} \mathrm{~erg/s}$ but with different opacity limits. Without opacity limit, $\varepsilon_{inj}=4 \times 10^{28} \mathrm{~erg/s}$ leads to runaway inflation, but with opacity limits runaway inflation disappears. Moreover, a lower limit corresponds to a stronger suppression on runaway inflation. Clearly, this numerical test indicates that runaway inflation is indeed caused by high opacity.

In the physical interpretation we use polytropic index $n=2.5$ near RCB to derive the power law for luminosity-temperature relation. Although equation of state is complicated in planet interior, we may apply the simple polytropic model $P\propto\rho^{1+1/n}$. Figure \ref{fig2e} shows pressure-density relation at three snapshots during evolution, i.e. before, at, and after runaway point. Near RCB (dots) $1+1/n=7/5$ ($n=2.5$) seems a good fit. In the interior ($\rho\gtrsim 0.1~{\rm g/cm^3}$) degeneracy $1+1/n=5/3$ ($n=1.5$) or partial degeneracy $1+1/n=2$ ($n=1$) seems a good fit.

To end this section, we estimate the tidal heating required for runaway inflation. For an eccentric orbit, tidal heating makes orbit circular (i.e. circularization),
\begin{equation}\label{dotE}
\dot{E}_{tide}\approx \frac{e^{2} G M_{*} M_{p}}{a \tau_{e}},
\end{equation}
where $a$ is semi-major axis, $e$ is eccentricity, and 
\begin{equation}\label{tau_e}
\tau_{e}\approx 0.33\left(\frac{Q_{p}^{\prime}}{10^{6}}\right)\left(\frac{M_{p}}{M_{\mathrm{J}}}\right)\left(\frac{M_{\odot}}{M_{*}}\right)^{3 / 2}\left(\frac{a}{0.04 \mathrm{AU}}\right)^{13 / 2}\left(\frac{R_{\mathrm{J}}}{R_{p}}\right)^{5} \mathrm{Gyr} ,
\end{equation}
is circularization timescale \citep{Goldreich1966}. Using the typical parameters of hot Jupiter, namely $M_*=M_\odot$, $M_p=M_J$, $a=0.04$ AU (i.e. orbital period $\approx 3$ days) and $R_p=2R_J$, we obtain $\tau_e\approx 1-10$ Myr for tidal $Q_p^{\prime}\approx 10^5-10^6$, and hence the critical rate of injected energy $\varepsilon_{cri}\approx 3.67\times10^{28}\mathrm{~erg/s}$ for runaway inflation corresponds to $e\approx 0.05-0.17$. Therefore, we can infer that such a hot Jupiter with $M_p\approx M_J$, $R_p\approx 2R_J$, $a\approx 0.04$ AU and $e>0.17$ has already experienced runaway inflation and does not exist, which is consistent with observations\footnote{https://exoplanetarchive.ipac.caltech.edu/}.

\section{Investigation on parameters}\label{sec:parameters}

In this section we investigate how the parameters influence runaway inflation. These parameters are planet mass $M_p$, core mass $M_{core}$, energy distribution $m_0$, and equilibrium temperature $T_{eq}$.

\begin{figure}
\centering
\subfigure[]{\includegraphics[width=0.32\textwidth]{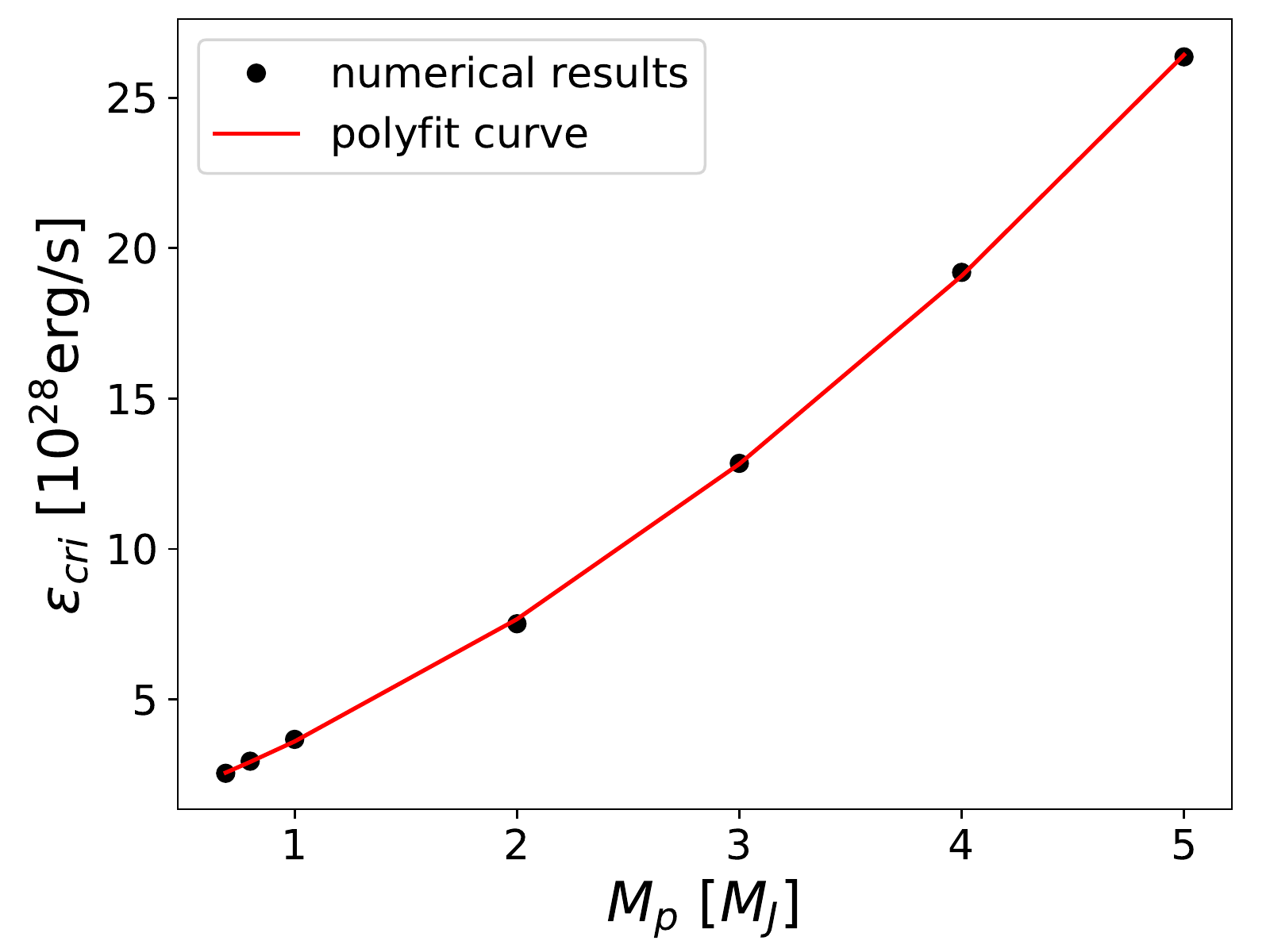}\label{fig3a}}
\subfigure[]{\includegraphics[width=0.32\textwidth]{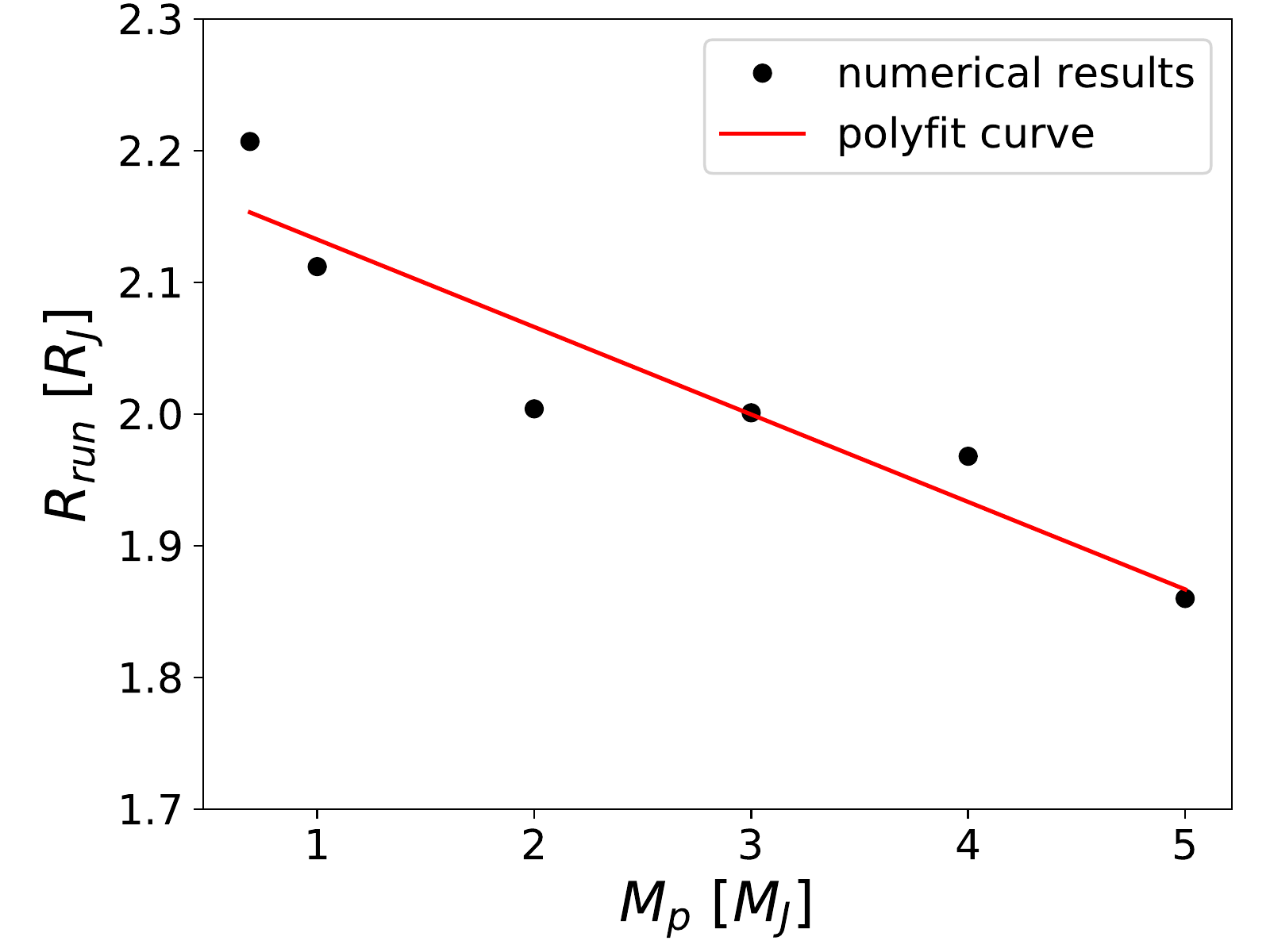}\label{fig3b}}
\subfigure[]{\includegraphics[width=0.32\textwidth]{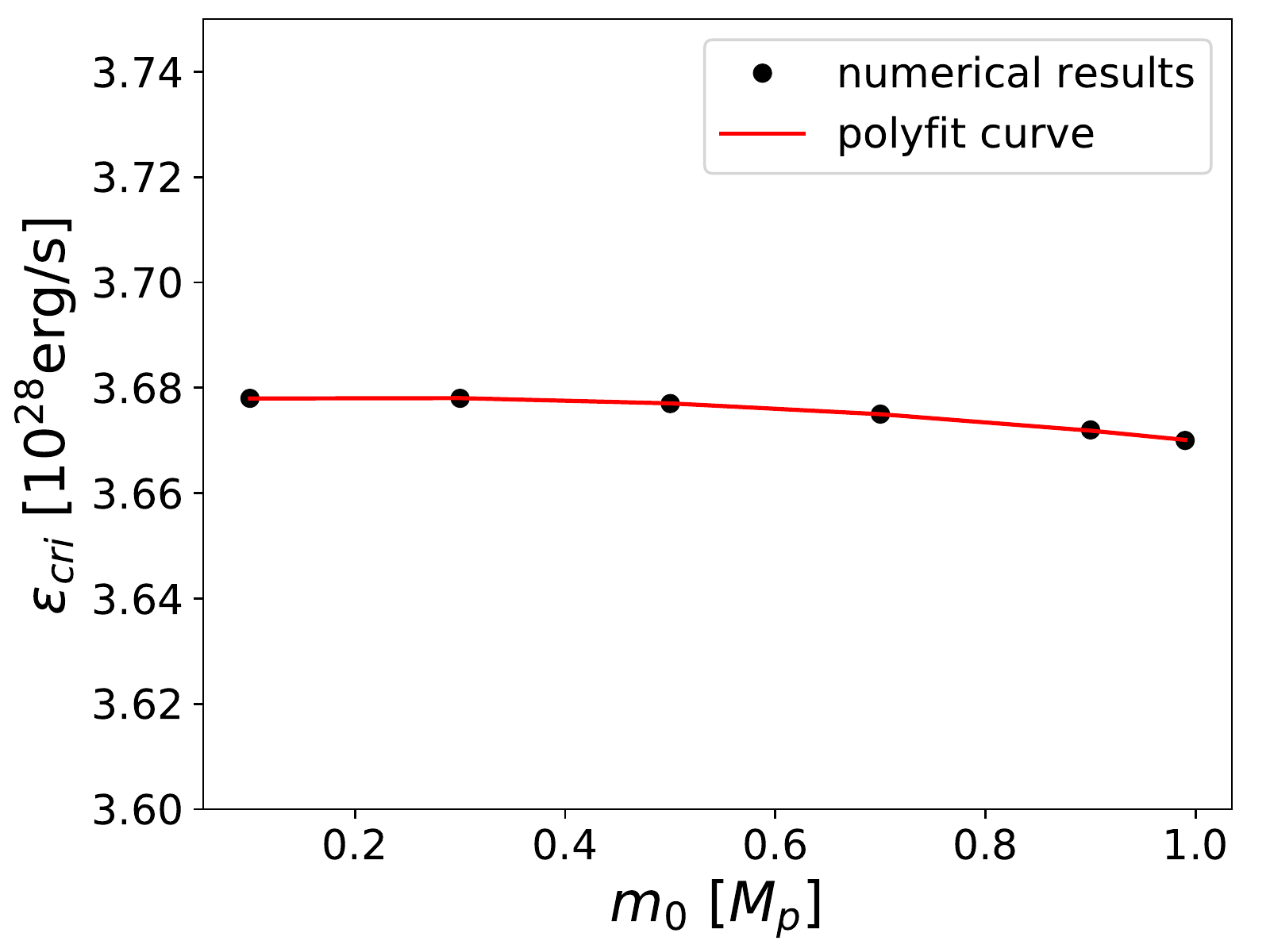}\label{fig3c}}
\subfigure[]{\includegraphics[width=0.32\textwidth]{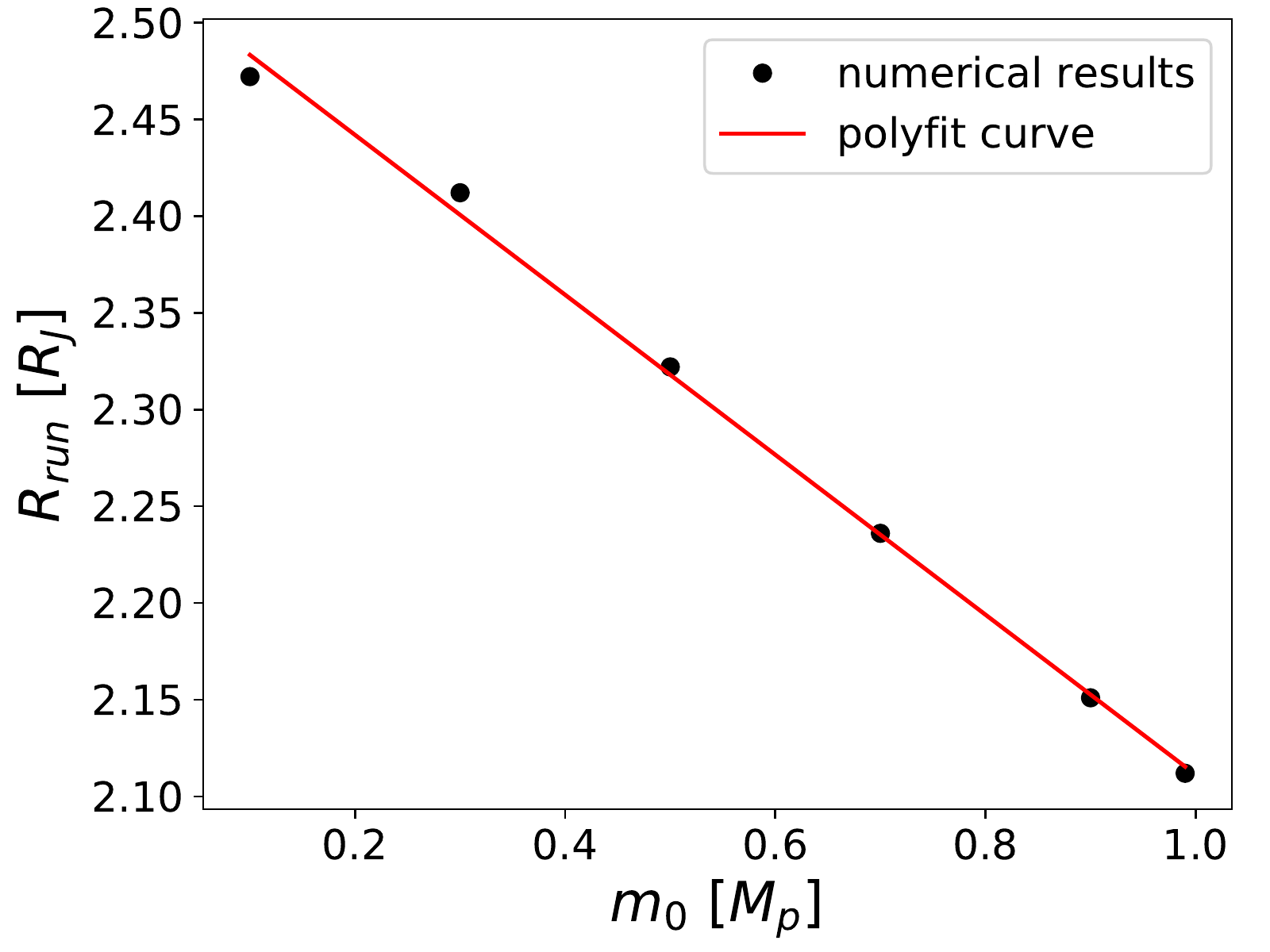}\label{fig3d}}
\subfigure[]{\includegraphics[width=0.32\textwidth]{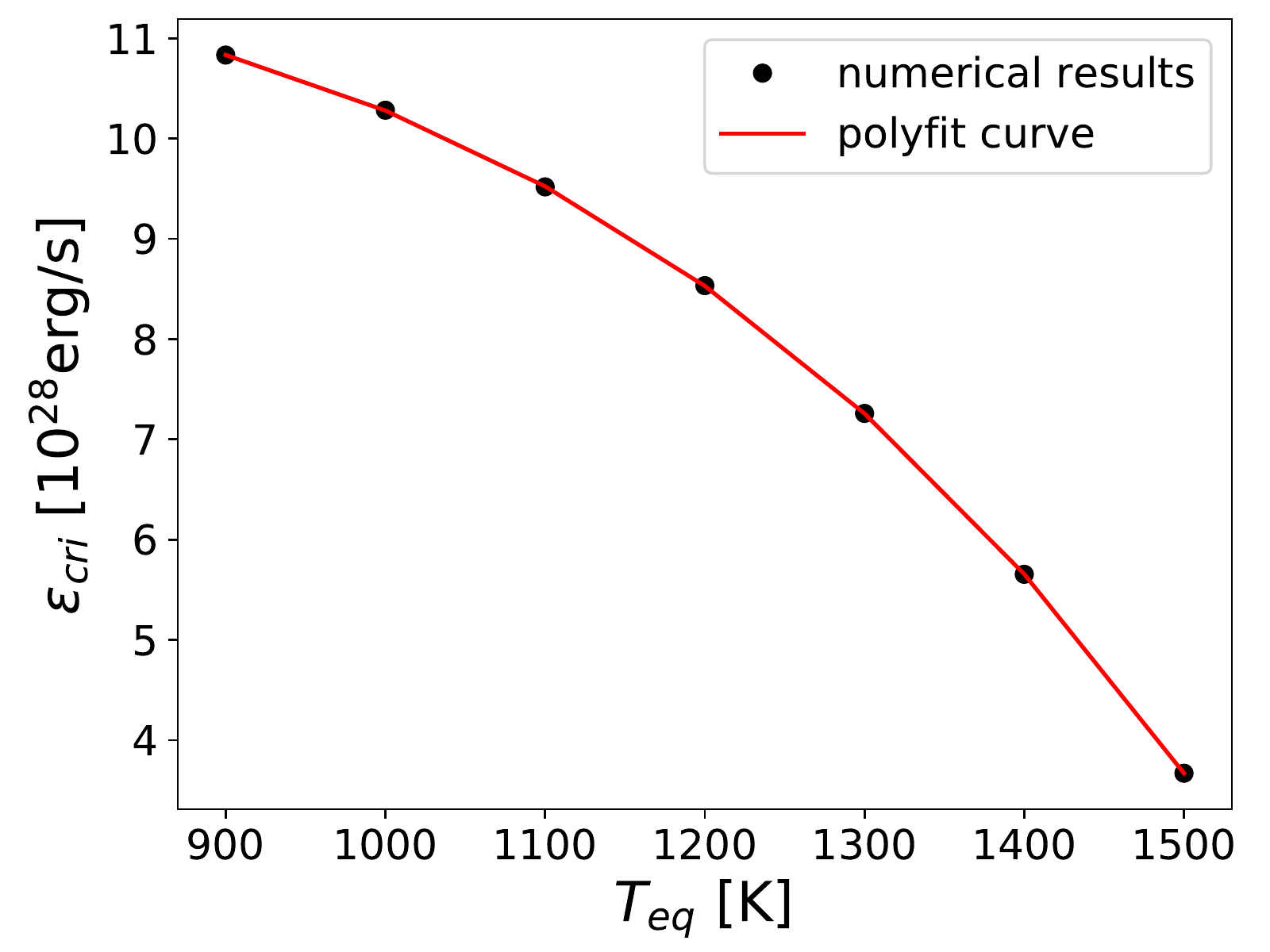}\label{fig3e}}
\subfigure[]{\includegraphics[width=0.32\textwidth]{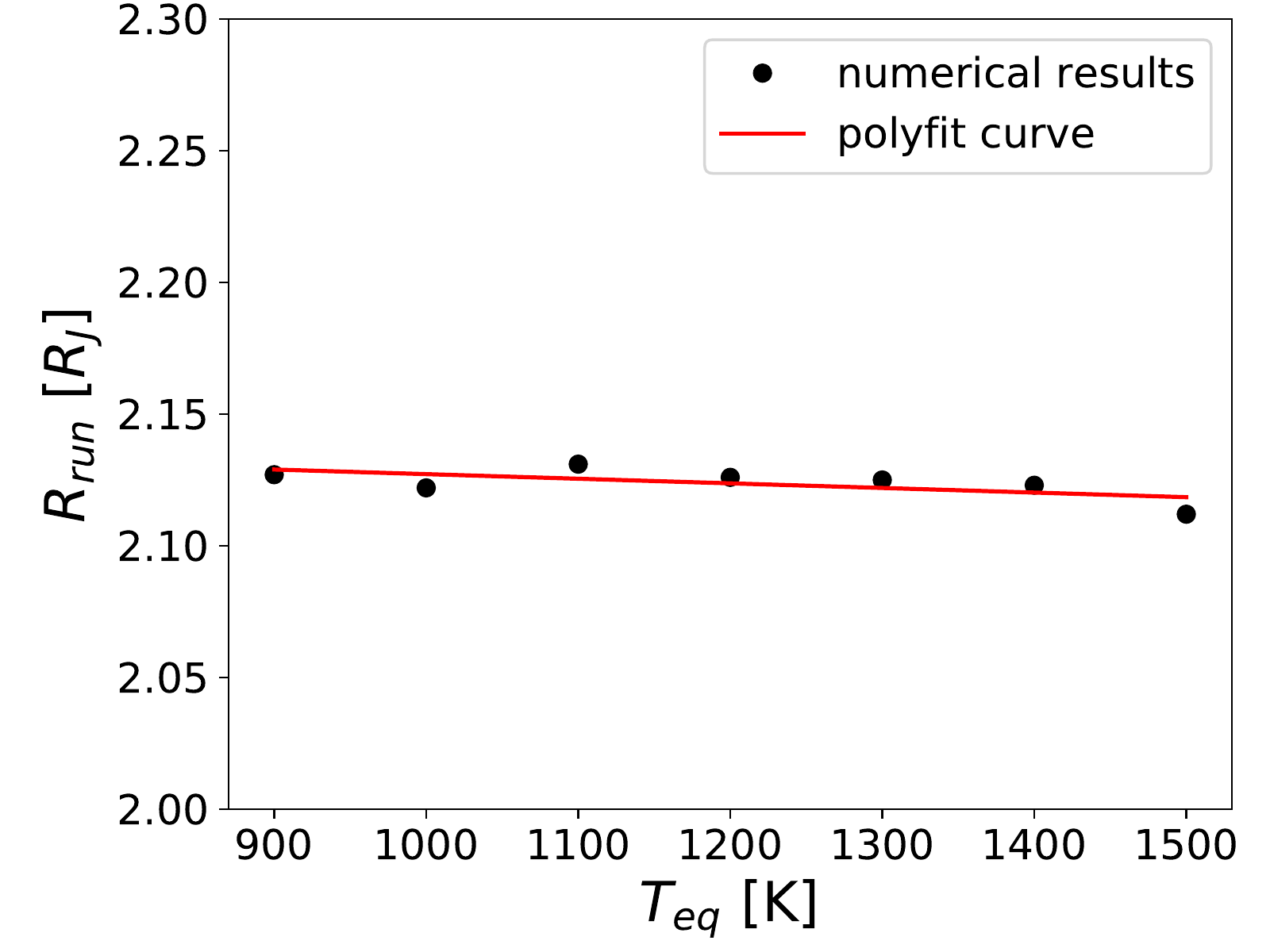}\label{fig3f}}
\caption{Investigation on parameters: (a) and (b) for planet mass, (c) and (d) for energy distribution, (e) and (f) for equilibrium temperature.}
\end{figure}

Firstly we investigate planet mass $M_p$. Figure \ref{fig3a} shows critical rate of injected energy $\varepsilon_{cri}$ required for runaway inflation versus planet mass. The polyfit index is about 1, which is physically reasonable because $\mathcal L_{intr}\propto M_p$. Since $\dot E_{tide}\le\varepsilon_{cri}\propto M_p$, by \eqref{dotE} and \eqref{tau_e} we find that the upper limit for eccentricity is $e\approx 0.17(M_p/M_J)^{1/2}$. Figure \ref{fig3b} shows that runaway radius $R_{run}$ is smaller for higher mass. This result gives an upper limit for hot Jupiter radius to be $2.2R_J$.

Next we investigate core mass $M_{core}$ from 0 (i.e. coreless planet) until 50 Earth mass. We find that both $\varepsilon_{cri}$ and $R_{run}$ for runaway inflation are independent of core mass. In our calculations planet core is a rigid boundary but not directly involved in the physical process in the interior. It has been found that Jupiter may have a dilute core \citep{Wahl2017}. A dilute core of exoplanet might influence runaway inflation, but we do not investigate this problem in this paper.

We now investigate energy distribution $m_0$. Figure \ref{fig3c} shows that $\varepsilon_{cri}$ is almost independent of energy distribution $m_0$, because it is the total injected energy that matters but not the energy distribution. Figure \ref{fig3d} shows that a deeper heat source leads to a larger $R_{run}$ (uniform distribution yields the same result as $m_0=0.1M_p$). This is because of temperature at RCB. With the different energy distributions the temperature at RCB is always about 3000 K (the accurate numerical value is 2750 K) because $\varepsilon_{cri}=3.67\times 10^{28}~{\rm erg/s}$ keeps constant. Thus, a deeper heat source leads to more delayed heat transfer because gas is more condensed in the deeper interior, which accordingly leads to a larger $R_{run}$. This result gives an upper limit for hot Jupiter radius to be $2.5R_J$ with very central energy source or uniform energy distribution. However, according to the observations, energy source near RCB is more plausible (see next section) and we believe the upper limit to be $2.2R_J$.

To end this section , we investigate equilibrium temperature $T_{eq}$ due to irradiation from host star. Irradiation takes place at planet surface close to RCB, so that higher $T_{eq}$ corresponds to lower luminosity near RCB which more easily induces runaway inflation. Figure \ref{fig3e} shows that $\varepsilon_{cri}$ indeed decreases with increasing $T_{eq}$. Figure \ref{fig3f} shows that $R_{run}$ is almost independent of $T_{eq}$.

\section{Comparison to the observations}\label{sec:observation}

\begin{figure}
\centering
\includegraphics[width=0.8\textwidth]{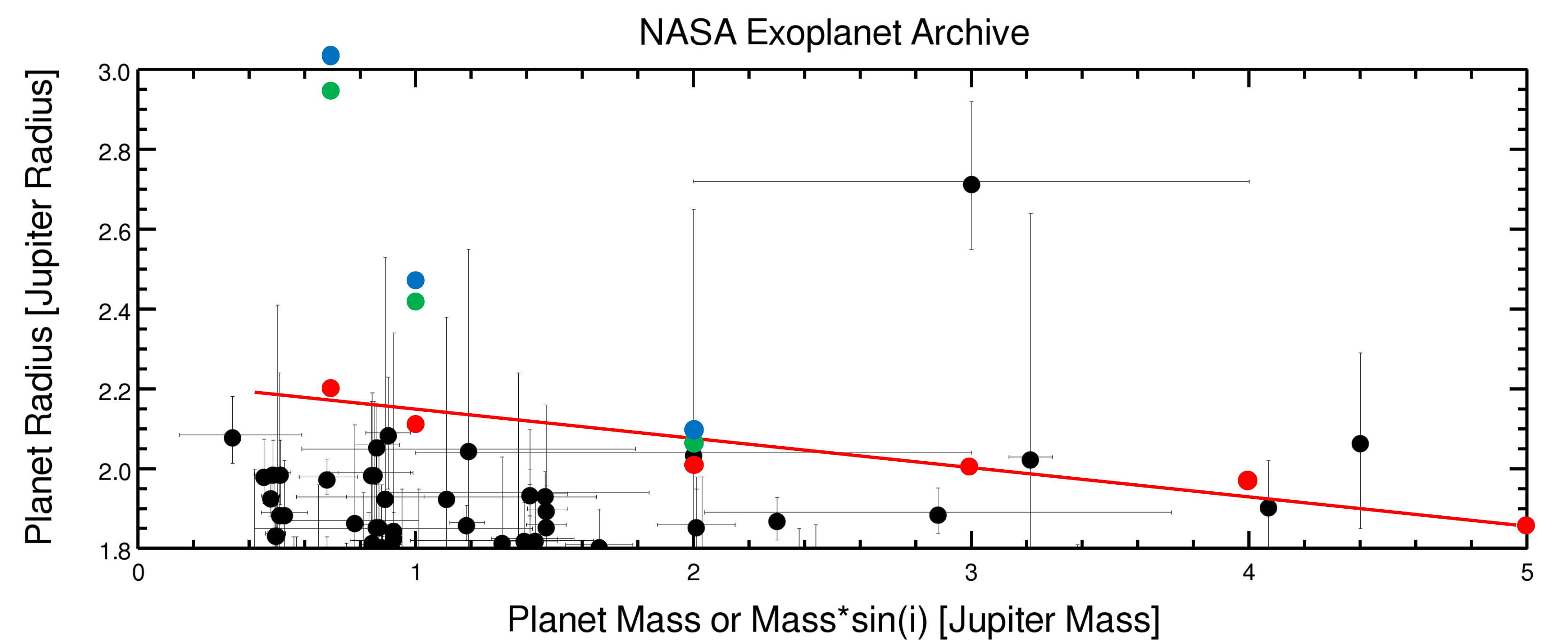}
\caption{The black dots denote the confirmed exoplanets with error bars (https://exoplanetarchive.ipac.caltech.edu/), the red dots denote our calculated $R_{run}$ with $m_0=0.99M_p$, the green dots with $m_0=0.3M_p$, and the blue dots with uniform distribution. Green and blues dots for $M_p\ge 3M_J$ are overlapped with red dots.}\label{fig4}
\end{figure}

Figure \ref{fig4} shows radius-mass relation of confirmed gaseous giant planets (black dots). The red line is a fitted boundary for radius-mass relation. Clearly, higher mass corresponds to smaller radius. The planet with $M_p\approx 3M_J$ and $R_p\approx 2.7R_J$ far away from the boundary line is a newly born planet PDS 70 b at age $\approx 5$ Myr when the protoplanetary disk has not been completely dispelled \citep{wang2020}, and the mechanism for its large radius must be not tidal heating since its orbital semi-major axis $\approx 20$ AU. On this figure, the red dots denote our calculated runaway radius with injected energy near RCB ($m_0=0.99M_p$), the green dots in the deep interior ($m_0=0.3M_p$), and the blue dots with uniform distribution (green and blues dots for $M_p\ge 3M_J$ are overlapped with red dots). The boundary line gives an upper limit for planet radius $2.2R_J$ which is consistent with our calculations. Moreover, the fact that the red dots fit very well with the observational boundary line indicates that tidal heating is very likely to locate near RCB but neither in the deep interior nor with uniform distribution.

\section{Summary and discussions}\label{sec:summary}

We numerically study planetary evolution with tidal heating and find the runaway inflation induced by high opacity near RCB. We then give the upper limit for hot Jupiter radius to be $2.2R_J$ (i.e. hot Jupiters can be large but not too large), and for eccentricity to be $0.17(M_p/M_J)^{1/2}$. Compared to the observations, we infer that tidal heating locates near RCB. 

Equations \eqref{dotE} and \eqref{tau_e} show that the tidal heating rate strongly depends on orbital parameters, and vice versa, the planet orbital dynamics certainly depends on the tidal heating since the latter alters planet radius and internal structure which in turn influences the orbital dynamics. In our study we did not consider this coupling, and it is worth studying in the future. Moreover, When runaway inflation occurs hot Jupiter loses its mass through atmosphere evaporation or accretion by host star, and how mass loss couples with orbital dynamics will be also a problem worth studying. In addition to tidal heating, as we have mentioned in \S\ref{sec:introduction}, Ohmic heating or inward transfer of irradiation can also induce radius inflation. Although this paper focuses on tidal heating, the mechanism of runaway inflation can be also applied to other heating sources. For example, when the electrically conducting atmosphere becomes thick, Ohmic heating may be more significant than tidal heating for inflation, and the observational constraints were studied with Bayesian analysis by \citet{thorngren2018}. In the situation that the heating is relevant to irradiation, the runaway inflation will be facilitated as planet radius increases, as studied again by \citet{thorngren2018}. These other sources are worth studying in the future.

\section*{Acknowledgements}
This project was initially illuminated by Douglas N. C. Lin, and Xianfei Zhang helps us on MESA. X. W. is financially supported by National Natural Science Foundation of China (grant no. 11872246 and 12041301) and Beijing Natural Science Foundation (grant no. 1202015).

\section*{Data availability}
The data underlying this article are available in the article.

\bibliographystyle{mnras}
\bibliography{paper}

\label{lastpage}
\end{document}